\documentclass[aps,a4paper,twocolumn,showpacs,preprintnumbers,amsmath,amssymb]{revtex4}

\usepackage{graphicx,amsmath}
\usepackage{dcolumn}
\usepackage{bm}
\usepackage{epsfig}
\usepackage{colordvi}

\newcommand{\lbfig}[1]{\refstepcounter{fig} \label{#1} }
\newcounter{fig}

\newcounter{One}
\newcounter{Two}
\newcounter{Three}
\def\putunder#1#2{\mathrel{
\setbox0=\hbox{#1}\setbox1=\hbox{\scriptsize #2}
\dimen0=-0.5\wd0 \advance\dimen0 by -0.5\wd1
\dimen1=0.5\wd0 \advance\dimen1 by -0.5\wd1
\hbox{\box0\kern\dimen0%
\vbox to 0pt {\hbox{\lower 0.7em \box1}\vss}%
\kern\dimen1}
}}
\def\beq{\begin{equation}}
\def\eeq{\end{equation}}
\def\bea{\begin{eqnarray}}
\def\eea{\end{eqnarray}}
\newcommand{\gsim}{\lower.7ex\hbox{$\;\stackrel{\textstyle>}{\sim}\;$}}
\newcommand{\lsim}{\lower.7ex\hbox{$\;\stackrel{\textstyle<}{\sim}\;$}}

\begin{document}

\title{Ramsey's Method of Separated Oscillating Fields and its Application to Gravitationally Induced Quantum Phaseshifts}

\author{H. Abele, T. Jenke, H. Leeb, J. Schmiedmayer}
\affiliation{%
Atominstitut der \"{O}sterreichischen Universit\"{a}ten\\
Stadionallee 2\\
1020 Wien, Austria
}%


\begin{abstract}

We propose to apply Ramsey's method of separated oscillating fields to the spectroscopy of the quantum states in the gravity potential above a vertical mirror. This method allows a precise measurement of quantum
mechanical phaseshifts of a Schr\"odinger wave packet bouncing off a
hard surface in the gravitational field of the earth. Measurements with ultra-cold neutrons will offer a sensitivity to Newton's law or hypothetical short-ranged interactions, which is about 21 orders of magnitude below the energy scale of electromagnetism.

\end{abstract}


\maketitle

\section{Introduction}

The system of a Schr\"odinger quantum particle with mass $m$
bouncing in a linear gravitational field is known as the quantum
bouncer~\cite{Gibbs,Rosu,qqb} and  "Quantum wave packet
revivals" can be found in \cite{Robinett}. Gravity tests with neutrons as quantum objects or within the classical limit are reviewed in~\cite{AbelePPNP}. Above a vertical mirror, the linear gravity potential leads to discrete energy eigenstates of a bouncing quantum particle. Such quantum states have been demonstrated at the Institut Laue-Langevin with ultra-cold neutrons in a previous collaboration~\cite{Nesvizhevsky1,Nesvizhevsky2,Nes05,Wes06}. Above a vertical mirror, linear gravity potential leads to discrete energy eigenstates of a bouncing particle. The lowest energy eigenvalues $E_n$, (n = 1, 2, 3, 4, 5), are 1.41 peV, 2.46 peV, 3.32 peV, 4.09 peV, and 4.78 peV. The energy levels together with the neutron density distribution are shown in Fig.~\ref{fig:1}. The idea of observing quantum effects in such a gravitational cavity was discussed with neutrons~\cite{Lushikov} or atoms~\cite{Wallis}.
\begin{figure}[htbp]
        \begin{center}
                \epsfig{file=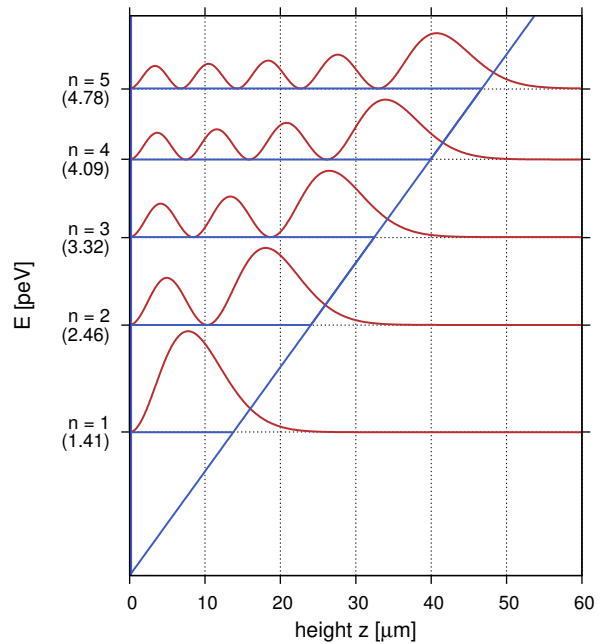}
                \lbfig{fig:1}
                \caption{Energy eigenvalues (blue) and neutron density distributions (red) for level one to five.}
        \end{center}
\end{figure}

An important feature of the quantum
bouncing ball -- in contrast to the harmonic oscillator problem --
is the fact that levels are not equidistant in energy. A
combination of any two states can therefore be treated as a
two-level system. The energy eigenstates in the gravity potential
can be coupled to a mechanical or magnetic oscillator field.
Transitions between quantum states in the gravitational field of the
earth i.e. a change of the state occupation can therefore be induced
similar to magnetic transitions, which occur when the oscillator
frequency equals one of the Bohr frequencies of the system. This
magnetic resonance method was in the original conception for
measurements of nuclear magnetic moments~\cite{Rab39,Kel39a}, but
soon it became a very general technique for radiofrequency
spectroscopy~\cite{Kel40}. Ramsey developed his method of separated
oscillating fields in which the oscillatory field is confined to a
region at the beginning and a region at the end with no oscillating
field in between~\cite{Ram56}. Variations of Ramsey's method is inherently
connected with precision measurements ranging from atomic clocks~\cite{A}
to atom in\-ter\-fero\-metry~\cite{B}, from NMR~\cite{C} to quantum-metrology~\cite{D}, or the related spin-echo technique~\cite{E}. That method has also been used to measure
the precession frequency of atoms, molecules or neutrons in a weak
magnetic field, for example in a search for permanent atomic or
neutron electric-dipole moments and in constructions of sensitive
magnetometers. The sensitivity is extremely high, because a quantum mechanical phase shift is converted into a frequency measurement. The sensitivity reached so far~\cite{Baker06} in a search for the
electric dipole moment of the neutron is 6.8 $\times$ 10$^{-22}$ eV, or one Bohr rotation within
6 days.

In analogy to these examples from electrodynamics, we discuss here an application
of Ramsey's method to probe the eigenstates in the gravity potential. Such a
technique should open a new way to precision gravity ex\-peri\-ments and
we propose to apply it to quantum states of neutrons or atoms in the gravitational field of the earth. This method will allow a precise measurements of
energy differences with a precision similar to the magnetic resonance technique. Here, we are sensitive to energy shifts of a
Schr\"odinger wave packet bouncing off a hard surface. Such energy shifts are expected from hypothetical gravity-like forces in the light of recent theo\-reti\-cal developments in higher-dimensional field theory and will allow searches for pseudo-scalar coupling of axions in the previously experimentally unaccessible astrophysical axion window~\cite{West,Bae07}, see Sec. III.

\section{Ramsey's method and its application to gravity potentials}
A quantum mechanical system that is described by two states can be understood
in analogy to a spin 1/2 system (assuming two states of a fictitious spin in
the multiplet, similarly to spin up and spin down states). The time
development of such systems is described by the Bloch equations.
In magnetic resonance
of a standard spin 1/2 system,
the energy splitting
results in the precession of the related magnetic moment in the magnetic
field. Transitions between the two states are driven by a transverse magnetic
radio frequency field. Similar concepts can be applied to any driven two level
system, e.g. in optical transitions with light fields. Here we apply this
picture to quantum states in the gravity field.

We start with a short description of Rabi's method~\cite{Rab39} to measure the
energy difference between a two-level system with a coupled oscillating field.
With $\omega_{pq}$, the frequency difference between the two states, $\omega$,
the frequency of the driving field, $\Omega_R$, the Rabi frequency and the
time $t$, the Hamiltonian $H$ is given by
\beq
H = \left( \begin{array}{cc}
\frac{\hbar\omega_{pq}}{2} & \frac{1}{2}\hbar\Omega_Re^{-i\omega t} \\
\frac{1}{2}\hbar\Omega_Re^{i\omega t} & \frac{-\hbar\omega_{pq}}{2}
\end{array} \right) \, .
\eeq

The probability of being found in the excited state as a function of
time is
\beq P(t) = \left(\frac{\Omega_R}{\Omega\acute{}_R}\right)^2
\sin^2\left(\frac{\Omega\acute{}_R}{2}t \right)\, ,
\label{Equ:Rabi}
\eeq
where the effective Rabi frequency is
\beq
\Omega\acute{}_R=\sqrt{\Omega^2_R+(\omega_{pq}-\omega)^2}=
\sqrt{\Omega^2_R+\delta^2}\, ,
\eeq
with detuning $\delta$ from resonance. The sinusoidal population transfer is
referred to as Rabi flopping. It has been proposed to measure the energy
levels of a neutron in the gravitational field of the Earth with this method
(GRANIT experiment\cite{Kreuz09,Pignol07}). The periodic drive is given by
neutrons moving through a spatially oscillating magnetic field created by
horizontal conducting wires.

As we will show below, one can drive transitions between quantum
states in gravity above the mirror by vibrating the mirror surface.

Lets consider the motion of ultracold neutrons in the gravitational field
above a mirror. We assume the gravitational force to act in -z-direction,
while the mirror is aligned with the xy-plane, vibrating with amplitude $a$ in
z-direction. The motion in x- and y-direction is free and
completely decouples from that in z-direction. It suffices therefore to consider the
time-dependent Schr\"odinger equation restricted to the z-direction
\beq
\left\{ -\frac{\hbar^2}{2m}\frac{\partial^2}{\partial z^2} + mgz
+V_0 \Theta (-z+a\sin \omega t)\right\}\Psi =
i\hbar\frac{\partial\Psi}{\partial t} \, .
\label{Heaviside}
\eeq
Here, $g$ is the acceleration of gravity, $m$ is the mass of the neutron and
$\Theta$ is the Heaviside step function. The potential $V_0\approx 100$~neV
associated with the substance of the mirror is repulsive and much larger than
eigenenergies of the lowest quantum states in the gravitational field. Therefore
Eq. (\ref{Heaviside}) must be solved with the boundary condition
$\Psi(z=a\sin\omega t,t)=0$. For further considerations it is preferable to
introduce  $\tilde z = z-a\sin \omega t$ and to transform
Eq. (\ref{Heaviside}) into the rest frame of the mirror,
\beq
\left\{ H_0 + W(\tilde z,t)\right\}\tilde\Psi =
i\hbar\frac{\partial\tilde\Psi}{\partial t}
\label{restframe}
\eeq
where
\begin{eqnarray}
H_0 & = &-\frac{\hbar^2}{2m}\frac{\partial^2}{\partial \tilde z^2} + mg\tilde z
+V_0 \Theta (-\tilde z)\, , \\
W(\tilde z,t) & = & a\left[ mg\sin\omega t+i\hbar\omega\cos\omega
  t\frac{\partial}{\partial\tilde z}\right]
\label{pert}
\end{eqnarray}
and $\tilde\Psi (\tilde z,t)=\Psi (z,t)$. The hamiltonian $H_0$ describes the
neutron in the gravitational field above a mirror at rest. The second term
$W(\tilde z,t)$ accounts for the vibration of the mirror.

The solution can be expressed in terms of the eigenfunctions $\psi_n(z)$ of
$H_0$

\beq
\Psi(z,t) =\sum_n C_n(t)e^{-iE_nt/\hbar}\psi_n(z)
\eeq

with time-dependent coefficients $C_n(t)$. Projection of Eq. (\ref{restframe})
on the eigenstates of $H_0$ yields a system of differential equations for the
coefficients $C_n(t)$
\beq
i\hbar \cdot \frac{d}{dt} C_n(t) = \sum_k \langle \psi_n | W | \psi_k
\rangle \cdot C_k(t) \cdot e^{i\omega_{nk}t}\, .
\eeq
The transitions between different quantum states is governed by the matrix
elements of $W(\tilde z,t)$ defined in (\ref{pert})
\beq
\langle\psi_n|W|\psi_k\rangle =a\left[ mg \delta_{n,k}\sin\omega t
+i\hbar\omega Q_{n,k} \cos\omega t \right]
\label{transmat}
\eeq
with
\beq
Q_{n,k}=\int_0^\infty dz \psi_n(z)\frac{d}{dz}\psi_k(z)\, .
\label{Qnk}
\eeq
The relevant overlap integrals $Q_{n,k}$ for the transitions between
the lowest eigenstates in the gravitational field are given in
Tab. \ref{tab:Qnk}.

\begin{table}
\begin{tabular}{|c|rrrrr|}
\hline
 & $k=1$ & $k=2$ & $k=3$ & $k=4$ & $k=5$ \\
\hline
$n=1$ &  0.00000 &  0.09742 & -0.05355 &  0.03831 & -0.03040 \\
$n=2$ & -0.09742 &  0.00000 &  0.11894 & -0.06314 &  0.04419 \\
$n=3$ &  0.05355 & -0.11894 &  0.00000 &  0.13458 & -0.07031 \\
$n=4$ & -0.03831 &  0.06314 & -0.13458 &  0.00000 &  0.14724 \\
$n=5$ &  0.03040 & -0.04419 &  0.07031 & -0.14724 &  0.00000 \\
\hline
\end{tabular}
\caption{Relevant overlap integrals $Q_{n,k}$ defined in Eq. (\ref{Qnk}) for
 the five lowest eigenstates in the gravitational field in $\mu$m$^{-1}$.}
\label{tab:Qnk}
\end{table}

The physics behind the transitions between the energy eigenstates of the
quantum bouncer caused by a vibrating mirror or an oscillating potential
is related to earlier studies of energy transfer when matter waves
bounce of a vibrating mirror~\cite{Hamilton87,Felber90,Felber96,Hils98} or
on a time dependent crystal~\cite{Steane95,Szriftgiser96,Bernet96}.  In the
later cases the transitions are between continuum states, in the quantum
bouncer between discrete eigenstates. Most interesting for our proposal to drive transitions between eigenstates of the quantum bouncer with a vibrating mirror is the physics of reflection of a neutron by an oscillating potential step as has been
investigated at the research reactors Munich and Geesthacht (FRM and
FRG)~\cite{Felber96}, however in a different energy regime.

Applying Ramsey's resonance method with separated oscillating fields will allow a careful
measurement of the energy eigenstates states of the quantum bouncer~\cite{AbeleETH08}. We propose to implement it with neutrons by traversing five regions as shown in
Fig.~\ref{fig:2}. The horizonal direction in space is considered as free
motion, while the vertical one is described by a one-dimensional time
dependent Schr\"odinger equation
(see e.g. Eq.~\ref{Heaviside}).

To implement Ramsey's method, one has to realize (1) a state selector,
(2) a region, where one applies a $\pi$/2 pulse creating the superposition of
the two states, whose energy difference should be measured,
(3) a region, where the phase evolves,
(4) a second region to read the relative phase by applying a second $\pi$/2
pulse, and finally
(5) a state detector.

In the following we will describe all these components as they are shown in
Fig.~\ref{fig:2}.
\begin{figure}[htbp]
        \begin{center}
                \epsfig{file=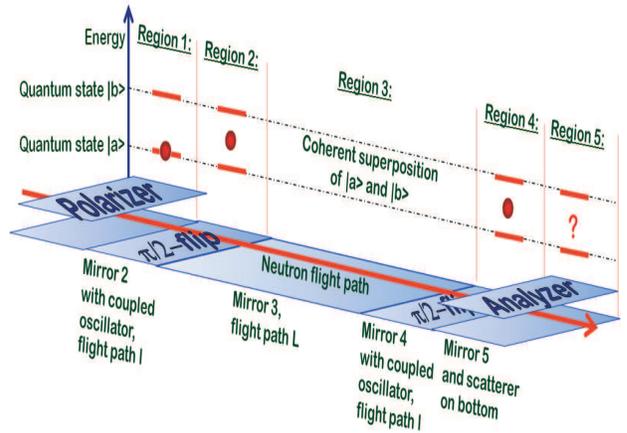,height=2.4in,width=3.2in}
                \lbfig{fig:2}
                \caption{Sketch of the proposal. Region 1: Preparation in a
                  specific quantum state, e.g. state one with
                  polarizer. Region 2: Application of first
                  $\pi$/2-flip. Region 3: Flight path with length $L$. Region
                  4: Application of second $\pi$/2-flip. Region 5: State
                  analyzer.}
        \end{center}
\end{figure}

In region one, neutrons are prepared in a specific quantum state $|p\rangle$
in the gravity potential following the procedure demonstrated in~\cite{Nesvizhevsky1}. A polished mirror on bottom and a rough absorbing
scatterer on top at a height of about 20 $\mu$m is a realization of a state
selector. It prepares neutrons into the ground state. Neutrons in higher,
unwanted states are scattered out of the system and absorbed i.e. $C_1$ = 1
and $C_n$ = 0 for n $>$ 1. A quantum mechanical description of such a system
can be found in~\cite{Wes06}. The neutron passage through a mirror-scatterer system has also been studied in a frame, where the rough scatterer surface has been treated as a time-dependent variation of the scatterer position~\cite{Voronin05}.

In region two of length $l$, the first of two identical oscillators is
installed. Here, transitions between quantum states $|p\rangle$ and
$|q\rangle$ are induced within a time $\tau$ according to
Equ.~\ref{Equ:Rabi}. The oscillator frequency at resonance for a transition
between states with energies $E_q$ and $E_p$ is \beq \omega_{pq} =
\frac{(E_q-E_p)}{\hbar}. \eeq The squared ratio of $\Omega_R$ and
$\Omega\acute{}_R$ as a function of the driving field $\omega$ for transitions
$|1\rangle$ $\rightarrow$ $|3\rangle$, $|1\rangle \rightarrow |2\rangle$
and $|2\rangle$ $\rightarrow$ $|3\rangle$ is shown in Fig.~\ref{fig:3}. For
the $|1\rangle \rightarrow |2 \rangle$ transition, which we chose as an
example for transitions in a two level system, $\omega_{12} = \omega_2 -
\omega_1$ = $2\pi \times 254$ s$^{-1}$.

\begin{figure}[htbp]
        \begin{center}
                \epsfig{file=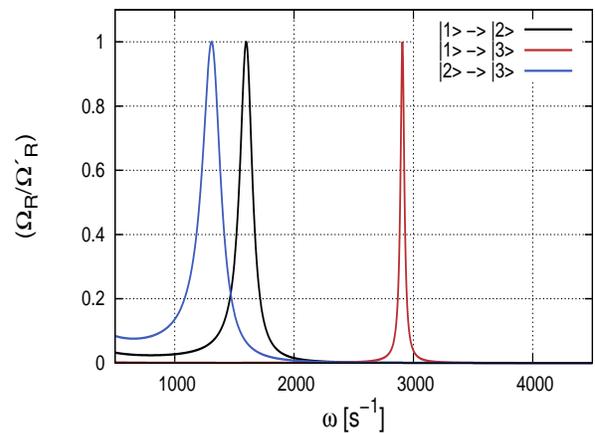,height=2.4in,width=3.2in}
                \lbfig{fig:3}
                \caption{Ratio $\left(\frac{\Omega_R}{\Omega\acute{}_R}\right)$ as a
function of $\omega$ [$s^{-1}$] for transition $|1>$$\rightarrow$
$|3>$, $|1>$$\rightarrow$ $|2>$ and $|2>$$\rightarrow$ $|3>$. The strength of the vibration was set to $1.0$  m/s$^2$.}
        \end{center}
\end{figure}
Driven on resonance ($\omega$ = $\omega_{pq}$), this oscillator drives the system into a coherent superposition of state $|p\rangle$ and $|q\rangle$.  A $\pi$/2-pulse, that is one with pulse area $\Omega_R\tau$ =
$\pi$/2, creates an equal superposition between state $|p\rangle$ and $|q\rangle$. This can be done by using oscillating magnetic gradient fields or by vibrating mirrors i.e. a modulation of the mirror potential in height.

In the intermediate region three, a non-oscillating mirror with a neutron flight
path of $L$ and flight time $T$ follows. It might be convenient to place a second mirror on top of the bottom mirror at a certain height $h$. It allows to tune the resonance frequency between $|a\rangle$ and $|b\rangle$ due to the additional potential, and it provides an effective doubling of sensitivity in a search for hypothetical axion induced
phase shifts or other fifth forces, see section~\ref{fifth}.

Subsequently, in region four a second
oscillator in phase with the oscillator in region two is placed. If the oscillating $\omega$ is equal $\omega_{pq}$ than the system is at resonance and we have a complete reversal of the state occupation between $|p\rangle$ and $|q\rangle$. There is no change in the relative phase of the oscillator and the quantum state of the neutron independent of the neutron velocity. In the other cases for $\omega\neq\omega_{pq}$, a velocity dependent relative phase shift builds up, since a slower neutron is in the region three longer and experiences a greater shift than a faster neutron.

Afterwards in section five, such a phase shift can be measured by transmission through a second state selector.

This method can be realized with some modifications to the previous setup in the following way: Neutrons are taken from the
ultracold neutron installation PF2 at ILL with a measured
horizontal velocity $v$ = 3.2 m/s $<$ $v$ $<$ 20 m/s. At the entrance of the
experiment, a collimator absorber system limits the transversal
velocity to an energy in the pico-eV range. The experiment itself is
mounted on a polished plane granite stone with an active and passive
antivibration table underneath. This stone is leveled using piezo
translators~\cite{Jenke08}. Inclinometers together with the piezo translators in a
closed loop circuit guarantee leveling with a precision better than
1 $\mu$rad~\cite{Stadler09}. A solid block with dimensions 10 cm $\times$ 15 cm
$\times$ 3 cm composed of optical glass serves as a mirror for
neutron reflection. The neutrons see a surface that is essentially
flat. In region one, an absorber/scatterer that is a rough mirror with a surface roughness of about 0.4 $\mu$m is placed above the first mirror at a height of 27 $\mu$m in order to select the first quantum state. The other states are efficiently removed, except for the second state, which is still present with a contribution of a few percent. In region two, a second mirror is placed
after the first one. Piezo elements attached underneath induce a fast modulation of the surface height with amplitude $a$ according to Eq.~\ref{Heaviside}.

As an example, we consider transitions between state $|1\rangle$ $\rightarrow$ $|2\rangle$ for the most probable velocity at the PF2/UCN beam position, 6 m/s. The length $l$ = 15 cm of this mirror is chosen in such a way to provide a neutron in a
superposition of these two quantum states after $\tau$ = 25.0 ms. Region three has a flight path of $L$ = 80 cm on a \-single mirror between the two oscillators in region two and, identical to region two, in region four. In region five, a state selector as an analyzer is placed, identical to the selector in region one but with a neutron detector behind for counting the transmitted neutrons. Calculated transition probabilities~\cite{Ram56} for $|p\rangle$ and $|q\rangle$ as a function of $\omega$ is shown in Fig.~\ref{fig:4} for different parts of the measured velocity spectrum.

This method can also be applied to stored ultra-cold neutrons. Fig.~\ref{fig:5} shows the theoretical Ramsey signal for a neutron storage time of 100 s. The appeal of a neutron storage lies in a very narrow resonance line.
\begin{figure}[htbp]
        \begin{center}
                \epsfig{file=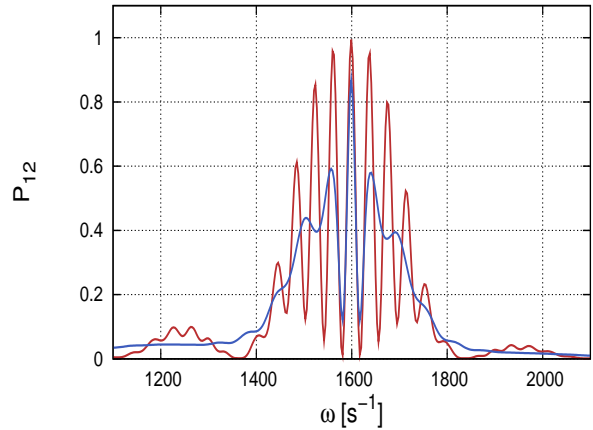,height=2.4in,width=3.2in}
                \lbfig{fig:4}
                \caption{Transition probability for a neutron velocity 5.5 m/s $<$ v
$<$ 6.5 m/s (red) and 3.0 m/s $<$ v $<$ 9.0 m/s (blue) as a function of $\omega$ [$s^{-1}$]. The strength of the vibration was set to $1.0 $m/s$^2$.}
        \end{center}
\end{figure}
\begin{figure}[htbp]
        \begin{center}
                \epsfig{file=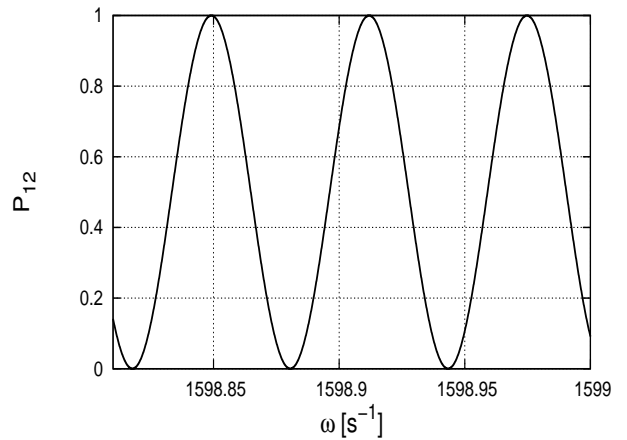,height=2.4in,width=3.2in}
                \lbfig{fig:5}
                \caption{Theoretical Ramsey signal for a neutron storage time of 100 s. Again, the strength of the vibration was set to $1.0 $m/s$^2$.}
        \end{center}
\end{figure}
A search for phase shifts are suggested in the next section.
\section{Phase shifts from hypothetical gravity-like fifth forces\label{fifth}}

Theoretical considerations arising from higher-dimensional gravity,
gauge forces or massive scalar fields suggest that the Newtonian
gravitational potential for masses $m_i$ and $m_j$ and distance $r$
should be replaced by a more general expression including a Yukawa
term,
\begin{equation}
        V(r) = -G\frac{m_i\cdot{m_j}}{r}(1-\alpha\cdot{e^{-r/\lambda}}),
        \label{yukawa}
\end{equation}
where $\lambda$ is the Yukawa distance over which the corresponding
force acts and $\alpha$ is a strength factor in units of Newtonian
gravity. $G$ is the gravitational constant. Most interesting, from
the experimental point of view, are scenarios, where the strength of
the new force is expected to be many orders of magnitude stronger
than Newtonian gravitation. Such forces are possible via abelian
gauge fields in the bulk~\cite{Arkani1,Arkani99,Antoniadis98,Antoniadis03}, (see also~\cite{stringADD,Koko04} for explicit
realizations in string theory). The strength of
the new force would be 10$^6$ $<$ $\alpha$ $<$ 10$^{12}$ stronger
than gravity, independent of the number of extra dimensions
$n$~\cite{Arkani99}. The observation of quantum states already tests speculations of this kind on
large extra dimensions of submillimeter size of
space-time~\cite{Abele,Nes,West}. Most recent theoretical developments support the
original proposal of large extra dimensions with bulk gauge fields
and more specific predictions for a high interaction strength can be
made. One proposal of Callin et al.~\cite{Callin} predicts
deviations from Newton's law on the micron scale on the basis of
supersymmetric large extra dimensions (SLED). The basic idea behind
this proposal is to modify gravity at small distances in such a way
as to explain the smallness of the observed cosmological constant.
The hope is to achieve this without changing non-gravitational
physics (which may be possible because of the small gravitational
response of the vacuum in specific models of SLED) and to link the
size of the extra dimensions to the energy density 10$^{10}$eV/m$^3$ governing the
observed dark energy component of the universe. In the concrete
constructions presently under discussion a radius $r$ of 10 microns as
well as the necessary interaction strength up to $\alpha$ = 10$^{6}$
may turn out to be well-motivated.

Other experimental limits on extra forces are derived from mechanical experiments and can be found, e.g., in~\cite{ad,Adel,Long,Chia,Fischb2,Bor,Geraci08}. Probing sub-micron forces by interferometry of Bose-Einstein condensated atoms has been proposed by~\cite{dim}. In practice, the experimental data are subject to corrections, which can be orders of magnitude larger than the effects actually searched for. It is therefore important to stress the completely different nature of possible systematic effects inherent to these micro-mechanical ex\-peri\-ments as compared to those in neutron experiments. In the former case gravitational interactions are studied in the presence of large van der Waals and Casimir-forces, which depend strongly on the geometry of the experiment, and the theoretical treatment of the Casimir effect is a difficult task. Currently, atomic force microscopes (AFM) measurements using
functionalized tips determine the limits on non-Newtonian
gravitation below  10 $~\mu$m. The best experimental data available
obtained with AFM, are claimed to be at the same level of accuracy
(1-2\%) as the numerical calculations of the Casimir force. The major obstacle for improvement in the theoretical
calculation is the fact that it is very hard to take the boundary
conditions of the tip and its functionalization properly into
account.

Our approach of probing Newtonian Gravity at the micron scale with
the help of Ramsey's Method of Se\-pa\-ra\-ted Oscillating Fields is advantageous
because of its small systematic effects. In contrast to atoms the electrical polarizability of neutrons~\cite{Schmiedmayer91} inducing such Casimir effects or van der Waals forces is extremely low. This together with its electric neutrality the neutron provides the key to a sensitivity of more than 10 orders of magnitude below the background strength of atoms.

The dynamics of such a quantum mechanical wave packet combines
quantum theory with aspects of Newtonian mechanics at short
distances. When a neutron with mass $m$ approaches the mirror, the
mass of this extended source might modify the earth acceleration
$g$, when strong non-Newtonian forces with range $\lambda$ and
strength $\alpha$ are present. For small neutron distances $z$ from
the mirror, say several micrometres, we consider the mirror as an
infinite half-space with mass density $\rho$. By replacing the
source mass $m_i$ by $dm_i$ and integrating over $dm_i$, the
modified Newtonian potential $\Delta V(z)$ is having the form
\begin{eqnarray}
\Delta V(z)&=&2\pi m \rho \alpha \lambda^2 G e^{-|z|/\lambda} \nonumber \\
&=& 8.47\times10^{-14}\, \alpha\, \lambda^2\, e^{-|z|/\lambda} \mathrm{peV}
\label{eqyuk}
\end{eqnarray}
with $\rho$ = 19 g/cm$^3$ (gold or tungsten coa\-ting) and $\lambda$ given in
$\mu$m. Taking these gravity-like forces into account, first order
perturbation theory predicts a shift of the $n$-th energy
eigenvalue~\cite{West},
\beq
\Delta E_n= \langle \psi_n |\Delta V(z)|\psi_n \rangle\, .
\eeq
They differ from state to state in the range of interest. The sensitivity can
been seen in Fig.~\ref{fig:6}, where the energy shift as a function of range
$\lambda$ is plotted for a fixed $\alpha$ = 10$^{12}$.
\begin{figure}
        \begin{center}
                \epsfig{file=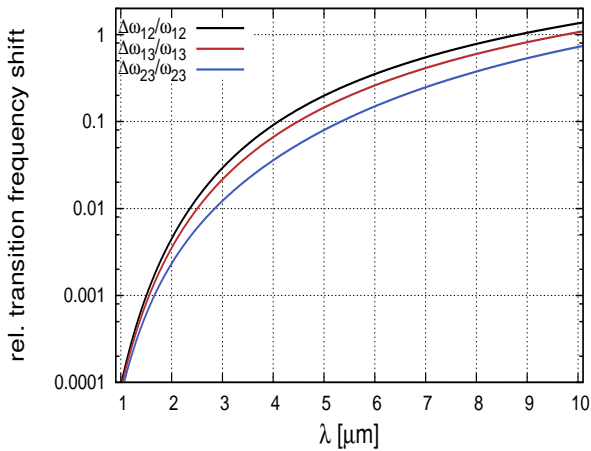,height=2.4in,width=3.2in}
                \lbfig{fig:6}
                \caption{The relative frequency shift vs. range $\lambda$. The strength of the modified Newtonian
                  potential $\Delta V(z)$ of Eq.~\ref{eqyuk} is set to $\alpha$ =
                  10$^{12}$.}
        \end{center}
\end{figure}

Newtonian gravity and hypothetical fifth forces evolve with
different phase information in the non-oscillating region. We expect
the following sensitivity for 50 days of beam time at the PF2-UCN beam position at the ILL:
With a count rate of 0.1 s$^{-1}$ for neutrons in the ground
state, we will have N = 430000 registered neutrons. Due
to the uncertainty principle $\Delta \Phi \Delta N$ $\geq$ 2$\pi$, we
estimate a minimal detectable phase shift of 9.6$\times$10$^{-3}$
radians. For an estimate of $T$ = 130 ms interrogation time (flight
path between the oscillators), the minimal resolvable energy shift
is \beq \Delta E = \Delta \Phi \hbar\ /T = 0.096\hbar/s =
4.8\times10^{-5} \mathrm{peV}. \eeq Together with Eq.~(\ref{eqyuk}), this
corresponds to a sensitivity of $\alpha = 7.4 \times 10^7,$
which is about three orders of magnitude better than existing limits right now.
In principle, ultracold neutrons can be stored and the time scale
can be increased to $T$ = 130 s. Therefore, a sensitivity to energy differences of $\Delta E$ = 4.8 $\times$ 10$^{-20}$ eV is feasible. This corresponds to an $\alpha = 7.6 \times 10^4$. With new neutron sources, which are under development right now, the source strength density is expected to be increased by two orders of magnitude. The statistical sensitivity of the new method is therefore around
 \beq \Delta E = 4.8\times10^{-21} \mathrm{eV}\eeq or \beq \alpha < 7.6 \times 10^{3}.\eeq
This is orders of magnitude better than existing limits right now.

Limits for hypothetical fifth forces can be easily interpreted as bounds of the strength of the matter couplings of axions. Axion interactions with a range within $20\,\mu m<\lambda<200\,mm$ (corresponding to axion masses
$10^{-6}\,\textrm{eV}<m_a<10^{-2}\,\textrm{eV}$), the 'axion
window', are still allowed by the otherwise stringent constraints
posed by cosmological data (see e.g.~\cite{kolb,rose}). The CP-violating spin-dependent part in presence of matter given by \cite{wilcz} is\beq V(\vec{r})=\hbar g_p g_s
\frac{\vec{\sigma}\cdot\vec{n}}{8\pi m c}\left(\frac{1}{\lambda
r}+\frac{1}{r^2}\right)\,e^{-r/\lambda}\;\;.\label{axpot1}\eeq
Here, $\vec{\sigma}$ denotes the neutron spin and $\vec{n}$ is a
unit vector related to the geometry of the macroscopic
matter configuration.

Integrating this potential over the geometry
of region three~\cite{West}, we arrive at $\lambda$ = 5 $\mu$m at a limit of
\beq \frac{g_s g_p}{\hbar
c}\lsim 5.3\cdot 10^{-23}\;\;.\label{axionlimit2}\eeq
for the dimensionless axion coupling strength. This is again eight orders of magnitude better than the only existing limit~\cite{West,Bae07} in the axion window from the previous experiment with neutrons.
\\

\section{Summary}

In conclusion, we discussed an application of Ramsey's method of oscillating fields to the quantum bouncer. It will allow high precision
spectroscopy of the energy eigenstates of a neutron bouncing on a flat vertical surface. Such Ramsey type interference measurements will
improve the sensitivity for neutron's coupling to gravity, to hypothetical short ranged forces or the influence of the cosmological constant. A sensitivity of 21 orders of magnitude below the strength of electromagnetism is found, when the energy $\Delta E$ = 4.8$\times$10$^{-21}$ eV of Eq. (17) is compared with the Rydberg energy of 13.6 eV, which is the energy scale of electromagentically bound quantum systems. Such an energy change corresponds to a strength $\alpha \sim 7.6 \times 10^3$ compared to gravity or to $\frac{g_s g_p}{\hbar
c}\sim 5.3\cdot 10^{-23}$, the axion coupling strength, at a range $\lambda$ = 5 $\mu$m.

The new method profits from small systematic effects in such systems, mainly due to the fact that in contrast to
atoms, the electrical polarizability of neutrons is extremely low.
Neutrons are not disturbed by short range electric forces such as
van der Waals or Casimir forces. Together with its neutrality, this
provides the key to a sensitivity of several orders of magnitude below the
strength of electromagnetism.

\section{Acknowledgment}

H.A. would like to thank D. Dubbers, R. Gaehler, and R. Golub for useful discussions.
This work has been supported by the German Research Foundation under Contract No. Ab128/2-1.

\end{document}